\documentclass{webofc}

\usepackage[varg]{txfonts} % Web of Conferences font
\usepackage{hyperref}
\usepackage{url}
\hypersetup{colorlinks=true,citecolor=blue,urlcolor=blue,linkcolor=blue}
\usepackage{graphicx}
\usepackage{subcaption}
\usepackage{pgfplots}
\pgfplotsset{compat=1.18}
\usetikzlibrary{arrows}
\usetikzlibrary{arrows.meta}
\usetikzlibrary{decorations.pathmorphing}
\usetikzlibrary{decorations.markings}

\newcommand{\pythia}{\textsc{Pythia}}

\begin{document}

\title{Interpreting the top quark mass using simulations of top-flavoured mesons}

\author{\firstname{Alexander} \lastname{Lind}\inst{1}\fnsep\thanks{\email{alexander.lind@roma1.infn.it}}
}

%\institute{INFN, Laboratori Nazionali di Frascati, Via E. Fermi 54, I-00044 Frascati (RM), Italy}
\institute{INFN, Sezione di Roma, Piazzale Aldo Moro 5, 00185 Roma, Italy}

\abstract{Active work is being done on the issue of the interpretation of the top quark mass in terms of well-defined theoretical definitions. This work considers this issue by simulating fictitious heavy-light top-flavoured mesons with a fixed mass in \textsc{Pythia}~8.3. Within Heavy Quark Effective Theory, the mass of the top mesons can be related to the pole mass of the top quark. By fitting final-state observables in top pair production at current hadron and future lepton colliders, we extract a negative mass difference between the input \textsc{Pythia} mass for the top quark and the pole mass of around $240$--$330$~MeV, consistent with $\mathcal{O}(\Lambda_{\rm QCD})$.}

\maketitle

%%%%%%%%%%%%%%%%%%%%%%%%%%%%%%%%%%%%%%%%%%%%%%%%%

\section{Introduction}
\label{sec:intro}

A precise determination of the top quark mass $m_t$ is relevant for several theoretical questions, including electroweak precision tests~\cite{deBlas:2022hdk}, electroweak-vacuum stability~\cite{Degrassi:2012ry,Domenech:2020yjf}, and models of Higgs inflation~\cite{Rodrigues:2023kiz}. 

Since the top mass is a free parameter of the Standard Model (SM), its numerical value must be extracted from collider experiments. The latest average PDG value is $(172.57 \pm 0.29)$~GeV~\cite{ParticleDataGroup:2024cfk}, which comes from so-called \emph{direct measurements}, i.e.\ the extraction of the top quark mass from the kinematics of $t\bar{t}$ events, such as the analyses in refs.~\cite{ATLAS:2025bpp, CMS:2024irj}. This mass extraction happens by comparison to parton shower Monte Carlo event generators such as \textsc{Pythia}~\cite{Bierlich:2022pfr} or \textsc{Herwig}~\cite{Bellm:2015jjp,Bewick:2023tfi}. Due to this fact, the extracted mass is referred to as the \emph{Monte Carlo mass} $m_t^{\rm MC}$. 

Active work has been done on the theoretical side, to relate the extracted Monte Carlo top quark mass to well-defined theoretical definitions for the top mass such as the \emph{pole mass} $m_t^{\rm pole}$ or the $\overline{\rm MS}$ scheme mass $\overline{m}_t(\overline{m}_t)$. The relation between the Monte Carlo mass and e.g.\ the pole mass can be written on a general form as
\begin{equation}
    m_t^{\rm MC} = m_t^{\rm pole} + \underbrace{\delta m}_{\substack{\text{Possible}\\\text{shift}}} \pm \underbrace{\Delta m}_{\text{Uncertainty}} \,,
\end{equation}
where $\delta m$ is a possible theoretical shift whereas $\Delta m$ is an uncertainty due to the modelling of non-perturbative QCD effects, namely hadronisation dynamics. Indeed, there are two competing views in the literature:
\begin{itemize}
    \item $\delta m \simeq 0$, i.e.\ that the Monte Carlo mass can essentially be identified as the pole mass, up to the uncertainty $\Delta m \sim \mathcal{O}(\Lambda_{\rm QCD})$~\cite{Nason:2016tiy, Nason:2017cxd}, 
    \item $\delta m \neq 0$, i.e.\ the Monte Carlo mass is fundamentally different from the pole mass, with estimates of the shift to be from $200$~MeV to $1$~GeV~\cite{Fleming:2007qr,Hoang:2008xm,Butenschoen:2016lpz,Hoang:2017kmk,Hoang:2018zrp}. 
\end{itemize}

The idea of the presented work is to consider the hadronisation of top quarks into fictitious top-flavoured mesons, e.g.\ $T^{+} = (t\bar{d})$, $T^{0} = (t\bar{u})$, $T^{-} = (\bar{t}d)$, etc., and then let them decay according to the spectator model. One can then reconstruct the $B$-lepton invariant mass $m_{B\ell}$, from which it is possible to get a relation between $m_t^{\rm MC}$ and $m_T$ by fitting $m_{B\ell}$. A relation between $m_T$ and $m_t^{\rm pole}$ can then be stated using Heavy Quark Effective Theory (HQET). Finally, using these relations we can obtain an estimate of $\Delta m$ from hadronisation systematics. 

These proceedings are based on the study presented in ref.~\cite{Corcella:2026tnu}, where a detailed account of the analysis can be found. Here, I present a concise summary of the main results discussed in the conference talk.

%%%%%%%%%%%%%%%%%%%%%%%%%%%%%%%%%%%%%%%%%%%%%%%%%

\section{Simulation of top-flavoured mesons}
\label{sec:topmesons}

While it is well-known that the top quark decays before hadronisation due to its large mass and short lifetime, it would nonetheless be interesting to have a Monte Carlo framework capable of producing heavy-light top mesons in order to shed light on the top mass interpretation issue. In this work, we have considered the production of top-anti-top pairs at current hadron and future lepton colliders, simulated with a modified version of the \textsc{Pythia} code that utilises the existing $R$-hadron machinery to produce top mesons with a fixed mass, $m_T$. For this purpose, we work in the narrow-width approximation by setting the top width to effectively zero. 

\begin{figure}[htb]
    \centering
    \begin{tikzpicture}
        \draw (0, 0) node[inner sep=0] {\includegraphics[width=0.7\textwidth,trim={5.8cm 6.5cm 9.5cm 7.0cm},clip]{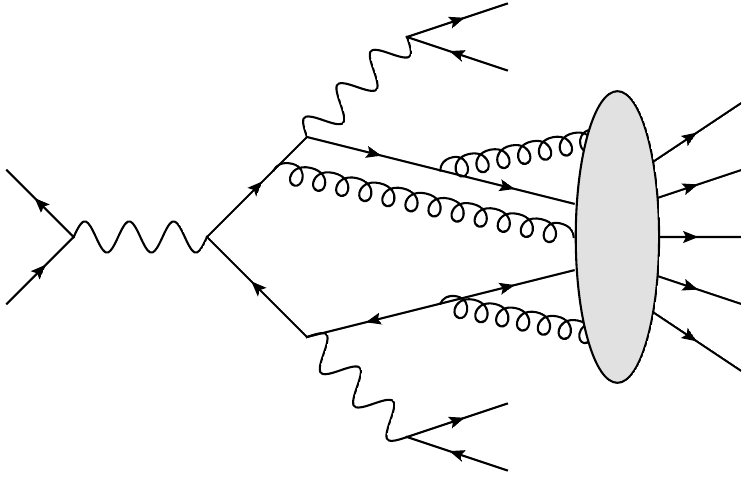}};
        \draw (-1.6, 0.9) node {$t$};
        \draw (-1.6, -0.9) node {$\bar{t}$};
        \draw (-4.6, 1.1) node {$e^{+}$};
        \draw (-4.6, -1.1) node {$e^{-}$};
        \draw (-0.5, 2.3) node {$W^{+}$};
        \draw (-0.5, -2.3) node {$W^{-}$};
        \draw (0.4, 1.3) node {$b$};
        \draw (0.4, -1.3) node {$\bar{b}$};
        \draw (2.0, 2.0) node {$\ell^{+}$};
        \draw (1.9, 2.9) node {$\nu$};
        \draw (2.0, -2.0) node {$\ell^{-}$};
        \draw (1.9, -2.9) node {$\bar{\nu}$};
        \draw (5.2, 0.0) node {\small Hadrons};
    \end{tikzpicture}
    \caption{Illustrative diagram for the standard top-anti-top pair production in $e^{+}e^{-}$ collisions, with leptonic decays, gluon emissions, and finally hadronisation. Reproduced from ref.~\cite{Corcella:2026tnu}.}
    \label{fig:feynmandiagramstandard}
\end{figure}

\begin{figure}[htb]
    \centering
    \begin{tikzpicture}
        \draw (0, 0) node[inner sep=0] {\includegraphics[width=0.6\textwidth,trim={5.8cm 3.7cm 6.8cm 3.6cm},clip]{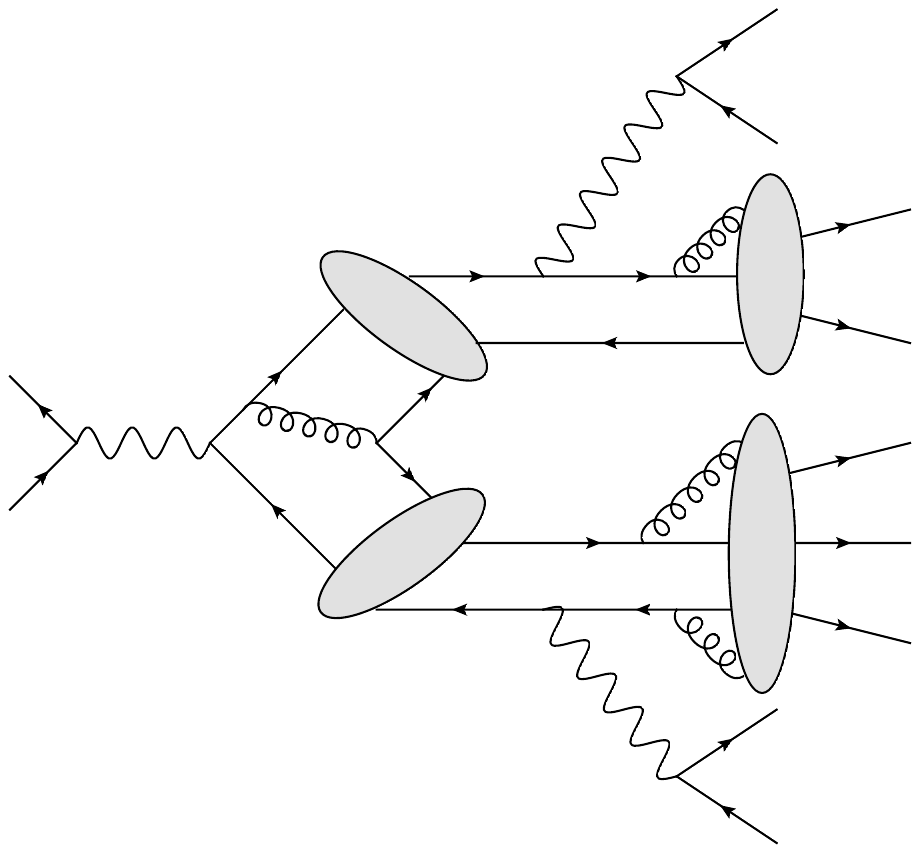}};
        \draw (-1.7, 0.8) node {$t$};
        \draw (-1.7, -1.0) node {$\bar{t}$};
        \draw (-1.5, 1.9) node {\small $T$-meson};
        \draw (-1.5, -2.03) node {\small $\bar{T}$-meson};
        \draw (-4.1, 0.7) node {$e^{+}$};
        \draw (-4.1, -1.0) node {$e^{-}$};
        \draw (0.9, 2.5) node {$W^{+}$};
        \draw (0.9, -2.6) node {$W^{-}$};
        \draw (1.7, 1.7) node {$b$};
        \draw (1.7, -1.9) node {$\bar{b}$};
        \draw (1.5, 0.5) node {\small Spectator};
        \draw (3.2, 2.6) node {$\ell^{+}$};
        \draw (3.1, 3.5) node {$\nu$};
        \draw (3.2, -2.5) node {$\ell^{-}$};
        \draw (3.1, -3.5) node {$\bar{\nu}$};
        \draw (4.8, 0.0) node {\small Hadrons};
    \end{tikzpicture}
    \caption{Illustrative diagram for the hadronisation of top quarks into top-flavoured mesons in $e^{+}e^{-}$ collisions, with decays according to the spectator model. Reproduced from ref.~\cite{Corcella:2026tnu}.}
    \label{fig:feynmandiagrammeson}
\end{figure}

An illustrative Feynman diagram for the process $e^{+}e^{-} \to t\bar{t}$ with leptonic decays of the $W$ bosons along with gluon emissions of the top and bottom quarks is shown in Figure~\ref{fig:feynmandiagramstandard}. Here top quarks undergo the usual decay with parton showers. Then the hadronisation is performed with a model such as the Lund string model~\cite{Andersson:1983ia} in the case of \textsc{Pythia}. Alternatively, Figure~\ref{fig:feynmandiagrammeson} illustrates an example of an event where the top quarks from the hard scattering hadronise into $T$-mesons which then decay according to a spectator model, where the resulting light spectator and $b$ quarks undergo the usual hadronisation. 

%%%%%%%%%%%%%%%%%%%%%%%%%%%%%%%%%%%%%%%%%%%%%%%%%

\subsection{Spectator decay model}
\label{sec:spectatordecay}

\begin{figure}[htb]
    \centering
    \begin{tikzpicture}
        \draw (0, 0) node[inner sep=0] {\includegraphics[width=0.399\textwidth,trim={9.0cm 9.8cm 9.5cm 5.9cm},clip]{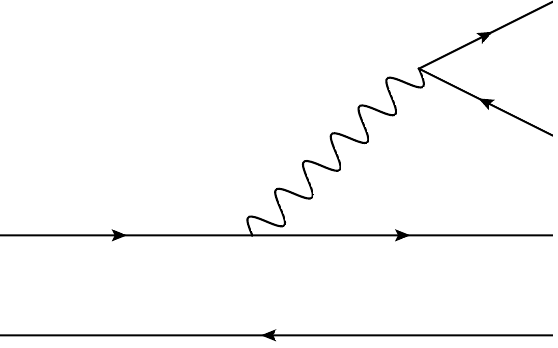}};
        \draw (-3.1, -0.6) node {$t$};
        \draw (-3.1, -1.6) node {$\bar{q}$};
        \draw (3.0, -0.6) node {$b$};
        \draw (3.0, -1.6) node {$\bar{q}$};
        \draw (0.0, 0.5) node {$W^{+}$};
        \draw (-3.8, -1.1) node {$T$};
        \draw (3.7, -1.1) node {$B$};
        \draw (3.1, 0.399) node {$\ell^{+}$};
        \draw (3.0, 1.7) node {$\nu$};
        % Left brace
        \draw (-3.399, -1.9) -- (-3.399, -0.3);
        \draw (-3.399, -0.3) -- (-3.25, -0.3);
        \draw (-3.399, -1.9) -- (-3.25, -1.9);
        % Right brace
        \draw (3.31, -1.9) -- (3.31, -0.3);
        \draw (3.31, -0.3) -- (3.16, -0.3);
        \draw (3.31, -1.9) -- (3.16, -1.9);
    \end{tikzpicture}
    \caption{Illustration of the spectator decay of a top-flavoured meson into a $B$-meson and a lepton-neutrino pair. Gluon emissions from the $b$-quark are not shown. Reproduced from ref.~\cite{Corcella:2026tnu}.}
    \label{fig:spectatordecay}
\end{figure}

The fictitious $T$-mesons will decay (similarly to $B$-mesons) according to the spectator decay model, where the $T$-meson is split into a top quark and a light spectator quark, with the top quark undergoing the usual leptonic $t \to Wb \to \ell\nu b$ decay,
\begin{equation}
    T^{+}(t\bar{d}) \rightarrow (b\bar{d})\ell^{+}\nu + X \,,
\end{equation}
where $X$ is mainly gluon radiation from the $b$ quark, and where $\ell = e$. This is illustrated in Figure~\ref{fig:spectatordecay} albeit without any gluon emissions from the $b$ quark. 

The light spectator quark will carry a fraction of the $T$-meson four-momentum,
\begin{equation}
    x_q = m_q / m_T \,,
\end{equation}
where $m_q$ is the mass of the spectator quark, which is set to $325$~MeV in \textsc{Pythia} by default. 

We note that the energetic $b$ quark produced in the $T$ decay subsequently initiates a sizeable parton shower. Therefore, as the spectator quark has very little effect on the decay dynamics as well as on the subsequent hadronisation process, it just contributes to the $T$-meson mass $m_T$.  

%%%%%%%%%%%%%%%%%%%%%%%%%%%%%%%%%%%%%%%%%%%%%%%%%

\subsection{Relating the top meson mass to the top pole mass with HQET}
\label{sec:hqet}

Following Heavy-Quark Effective Theory (HQET)~\cite{Neubert:1996si,Manohar:2000dt}, the mass of a heavy-light meson $Q$ containing a heavy quark $q$ can in general be expanded as
\begin{equation}
    m_Q = m_q + \overline{\Lambda} - \frac{\Delta m^2}{2 m_q} + \mathcal{O}\left ( \frac{1}{m_q^2} \right ) \,,
\end{equation}
where
\begin{itemize}
    \item $\overline{\Lambda}$ comes from terms in the Lagrangian independent of $m_q$,
    \item $\Delta m^2$ comes from $\mathcal{O}(1/m_q)$ terms in the effective HQET Lagrangian,
    \begin{equation}
        \Delta m^2 = \lambda_1 - 2 \left [ J (J + 1) - \frac{3}{2} \right ] \lambda_2(m_q) = \lambda_1 + n \lambda_2(m_q) \,,
    \end{equation}
    \item and $J$ is the total meson spin.
\end{itemize}
Here $\bar{\Lambda}$, $\lambda_1$, and $\lambda_2$ are scheme-dependent universal QCD parameters. Exploiting flavour-independence and subtracting $m_B$ and $m_T$ we get
\begin{equation}
    m_T = m_t + m_B - m_b - \lambda_1\left(\frac{1}{2m_t}-
    \frac{1}{2 m_b}\right) - \frac{n}{2} \left [ \frac{\lambda_2(m_t)}{m_t} - \frac{\lambda_2(m_b)}{m_b} \right ] \,,
\end{equation}
which, due to the large top mass, can be approximated by a constant shift between the meson and quark masses,
\begin{equation}
    m_T \approx m_t + \bar{\Lambda} \,,
\end{equation}
where $\bar\Lambda$ can be evaluated as
\begin{equation*} 
    \bar\Lambda = m_t + m_B - m_b + \frac{1}{2 m_b} \big [ \lambda_1 + n \, \lambda_2(m_b) \big ] \,.
\end{equation*}
We set $n = +3$ and use the PDG values for $m_t$, $m_b$, and $m_B$ in the pole mass scheme~\cite{ParticleDataGroup:2024cfk}. Generally we have
\begin{equation*}
    \vert \lambda_1 \vert \sim \lambda_2 \sim \Lambda_{\rm QCD}^2 \sim 0.1~\mathrm{GeV}^2 \,,
\end{equation*}
but a more precise determination is obtained from ref.~\cite{Jeong:1998gj} for $B$-meson decays give
\begin{equation*}
    \lambda_1 = (-0.58 \pm 0.23)~\mathrm{GeV}^2 \,,
\end{equation*}
while considerations in ref.~\cite{Nefediev:2024mjk} from lattice data give the relation
\begin{equation*}
    \lambda_2(m_b) \approx \frac{m_b}{2} \left ( m_{B^{\ast}} - m_B \right ) \approx (0.1080 \pm 0.0014)~\mathrm{GeV}^2 \,.
\end{equation*}
Finally, we obtain an approximate value of
\begin{equation*}
    \bar{\Lambda} = (0.473 \pm 0.064)~\mathrm{GeV} \,, \label{eq:lambdabar}
\end{equation*}
which agrees with the value from ref.~\cite{Nefediev:2024mjk} of $\bar{\Lambda} = (0.49 \pm 0.08)~\mathrm{GeV}$.

%%%%%%%%%%%%%%%%%%%%%%%%%%%%%%%%%%%%%%%%%%%%%%%%%

\section{Results}
\label{sec:results}

In this section we present results from our Monte Carlo study with \textsc{Pythia}~8.317. We have simulated standard $t\bar{t}$ production (referred to as \emph{standard samples}) as well as the hadronisation of top quarks into $T$-mesons as outlined in section~\ref{sec:topmesons} (referred to as \emph{$T$-meson samples}). In either type of samples, we vary the input top quark/meson mass $m_t^{\textsc{Pythia}}$ and $m_T$ respectively. Additionally we generate all results for two different collider setups: 
\begin{itemize}
    \item The current LHC Run 3 setup, $pp \to t\bar{t}$, $\sqrt{s} = 13.6$~TeV, and
    \item a potential future lepton collider (FLC), $e^{+}e^{-} \to t\bar{t}$, $\sqrt{s} = 1$~TeV. 
\end{itemize}

%%%%%%%%%%%%%%%%%%%%%%%%%%%%%%%%%%%%%%%%%%%%%%%%%

\subsection{\texorpdfstring{$m_{B\ell}$}{mBl} distributions and extraction of the mass shift}
\label{sec:mbl}

In this section we present results for the differential distribution in the invariant mass $m_{B\ell}$ of the $B$-hadron and the lepton from top quark and $T$-meson decays. 

\begin{figure}[htb]
    \centering
    \begin{subfigure}{.5\textwidth}
        \centering
        \includegraphics[width=\linewidth]{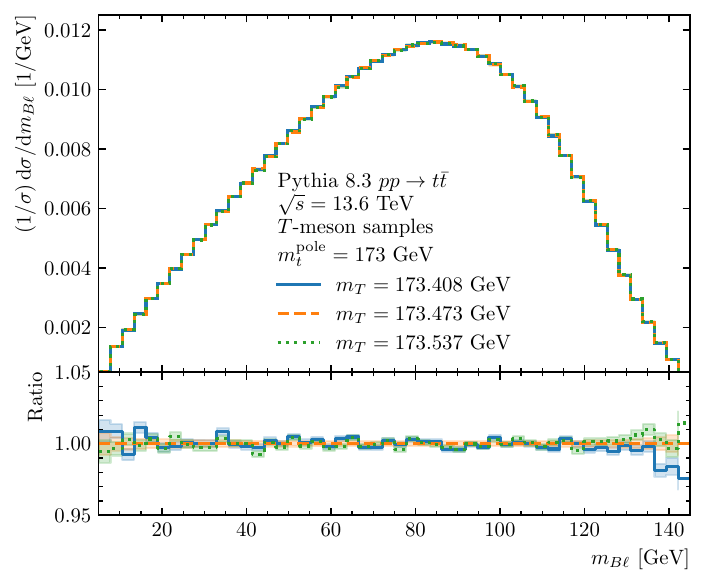}
        \caption{$pp \to t\bar{t}$, at $\sqrt{s} = 13.6$~TeV.}
        \label{fig:mblpp}
    \end{subfigure}%
    \begin{subfigure}{.5\textwidth}
        \centering
        \includegraphics[width=\linewidth]{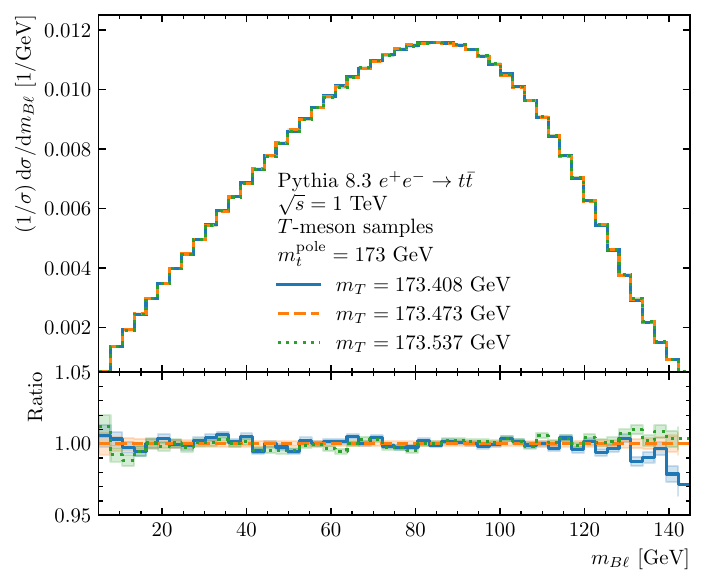}
        \caption{$e^{+}e^{-} \to t\bar{t}$, at $\sqrt{s} = 1$~TeV.}
        \label{fig:mblee}
    \end{subfigure}
    \caption{Differential distributions in $m_{B\ell}$ for the $T$-meson samples with fixed $m_{T} = m_t^{\rm pole} + \bar{\Lambda}$ and $m_t^{\rm pole} = 173$~GeV at the LHC (a) and FLC (b). The samples come from varying $m_T$ within its uncertainty in eq.~(\ref{eq:lambdabar}). Reproduced from ref.~\cite{Corcella:2026tnu}.}
    \label{fig:mbl}
\end{figure}

Figure~\ref{fig:mbl} shows the effect that the uncertainty on $\bar{\Lambda}$ in eq.~\ref{eq:lambdabar} has on the $m_{B\ell}$ distribution. Compared to the usual Monte Carlo error it can be considered negligible.

\begin{figure}[htb]
    \centering
    \begin{subfigure}{.5\textwidth}
        \centering
        \includegraphics[width=\linewidth]{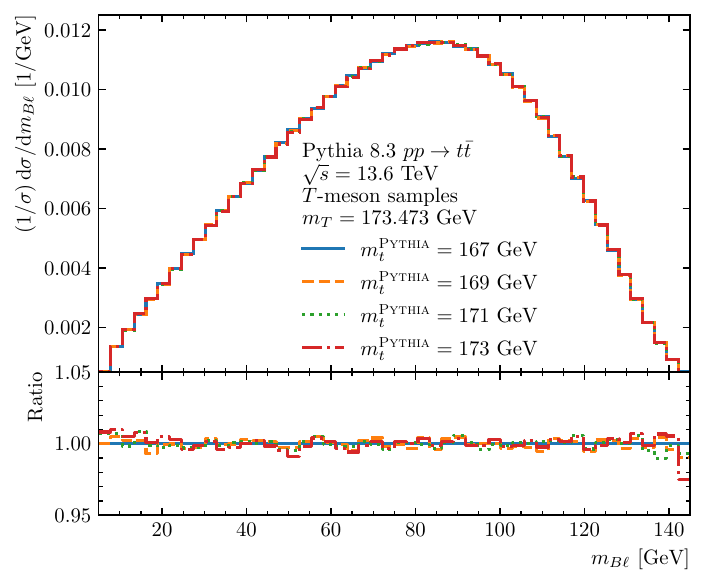}
        \caption{$pp \to t\bar{t}$, at $\sqrt{s} = 13.6$~TeV.}
        \label{fig:mblindependentpp}
    \end{subfigure}%
    \begin{subfigure}{.5\textwidth}
        \centering
        \includegraphics[width=\linewidth]{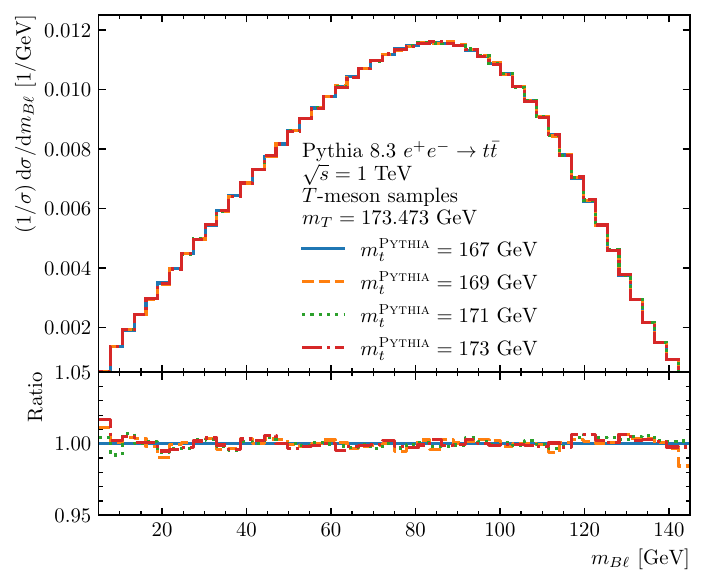}
        \caption{$e^{+}e^{-} \to t\bar{t}$, at $\sqrt{s} = 1$~TeV.}
        \label{fig:mblindependentee}
    \end{subfigure}
    \caption{Differential distributions in $m_{B\ell}$ for the $T$-meson samples for fixed $m_{T} = m_t^{\rm pole} + \bar{\Lambda}$ with $m_t^{\rm pole} = 173$~GeV and different choices of the input top mass $m_t^{\pythia{}}$. Reproduced from ref.~\cite{Corcella:2026tnu}.}
    \label{fig:mblindependent}
\end{figure}

Figure~\ref{fig:mblindependent} shows a validation plot demonstrating the independence of the $m_{B\ell}$ distribution for the $T$-meson samples on the input top quark mass in \textsc{Pythia}, $m_t^{\textsc{Pythia}}$. 

\begin{figure}[htb]
    \centering
    \begin{subfigure}{.5\textwidth}
        \centering
        \includegraphics[width=\linewidth]{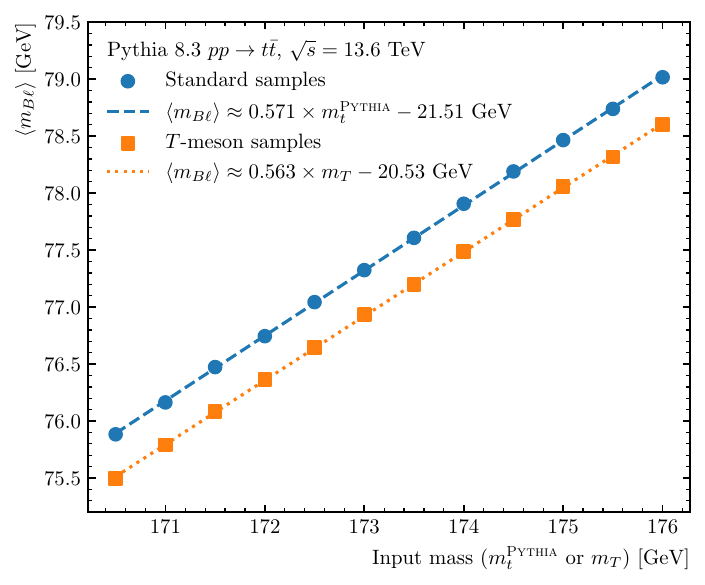}
        \caption{$pp \to t\bar{t}$, at $\sqrt{s} = 13.6$~TeV.}
        \label{fig:averagemblpp}
    \end{subfigure}%
    \begin{subfigure}{.5\textwidth}
        \centering
        \includegraphics[width=\linewidth]{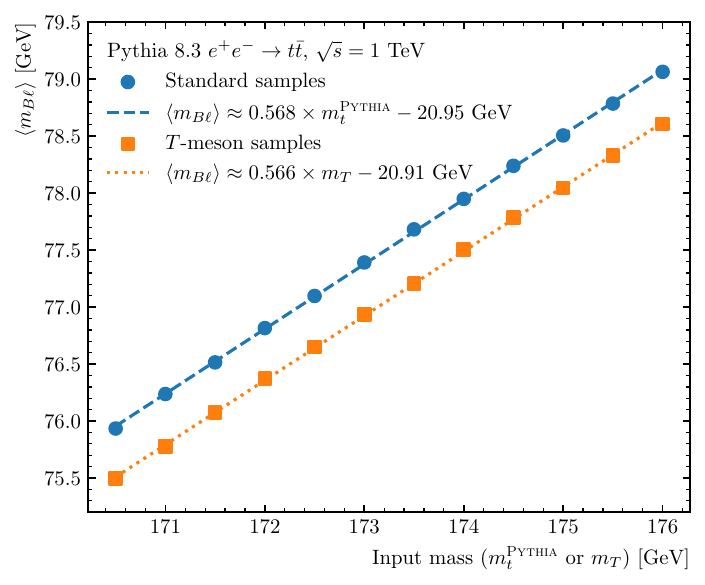}
        \caption{$e^{+}e^{-} \to t\bar{t}$, at $\sqrt{s} = 1$~TeV.}
        \label{fig:averagemblee}
    \end{subfigure}
    \caption{Linear fits of the average $\langle m_{B\ell} \rangle$ for the $T$-meson samples (in orange) and the standard $t\bar{t}$ samples (in blue) varying $m_T$ and $m_t^{\pythia{}}$ respectively. Reproduced from ref.~\cite{Corcella:2026tnu}.}
    \label{fig:averagemblboth}
\end{figure}

An observable related to the $m_{B\ell}$ distribution is the event-wise average $\langle m_{B\ell} \rangle$ which can be expressed with a linear dependence on the input \textsc{Pythia} top mass for the standard samples,
\begin{equation}
    \langle m_{B\ell} \rangle = a \times m_t^{\rm Pythia} + b \,,
\end{equation}
and on the top meson mass in $T$-meson samples,
\begin{equation}
    \langle m_{B\ell} \rangle = a' \times m_T + b' \,.
\end{equation}
The linear fits are shown in Figure~\ref{fig:averagemblboth} and give for the standard samples,
\begin{equation}
    \langle m_{B\ell} \rangle_{pp} \simeq 0.571\  m_{t}^{\rm Pythia} -21.51~{\rm GeV} \,,
  \ \   \delta=0.011~{\rm GeV}\,;
\end{equation}
\begin{equation}
    \langle m_{B\ell} \rangle_{ee} \simeq 0.568\  m_{t}^{\rm Pythia} -20.95~{\rm GeV} \,,
   \ \  \delta=0.014~{\rm GeV}\,.
\end{equation}
The linear fits for the $T$-meson samples give
\begin{equation}
    \langle m_{B\ell} \rangle_{pp} \simeq 0.563\  m_T -20.53~{\rm GeV} \,,
  \ \   \delta=0.010~{\rm GeV}\,;
\end{equation}
\begin{equation}
    \langle m_{B\ell} \rangle_{ee} \simeq 0.566\  m_T -20.91~{\rm GeV} \,,
   \ \  \delta=0.013~{\rm GeV}\,.
\end{equation}
The residual standard deviation of the fit is defined as
\begin{equation}
    \delta_{\rm fit} = \sqrt{\frac{1}{N-2} \sum_{i=1}^{N} \left [ \langle m_{B\ell} \rangle_i - \left ( a \times m_{t,i}^{\rm Pythia} + b \right ) \right ]^2} \,.
\end{equation}

The slopes agree well within the errors, i.e.\ $a \approx a'$. Re-performing the linear fits with fixed slopes, and using the HQET result, $m_T \approx m_t^{\rm pole} + \bar{\Lambda}$, with $\bar{\Lambda} \approx 0.473$~GeV, we find a mass difference of
\begin{eqnarray}
    m_t^{\rm Pythia} - m_t^{\rm pole} = -0.24 \pm 0.07~\text{GeV} \hspace{7mm} \text{(for $pp$ collisions),}
\end{eqnarray}
\begin{eqnarray}
    m_t^{\rm Pythia} - m_t^{\rm pole} = -0.33 \pm 0.07~\text{GeV} \hspace{7mm} \text{(for $e^{+}e^{-}$ collisions),}
\end{eqnarray}
which are consistent with a mass difference of $\mathcal{O}(\Lambda_{\rm QCD})$.

%%%%%%%%%%%%%%%%%%%%%%%%%%%%%%%%%%%%%%%%%%%%%%%%%

\subsection{Sensitivity to the recoil treatment in top decays}
\label{sec:sensitivity}

\begin{figure}[htb]
    \centering
    \begin{tikzpicture}
        \draw (0, 0) node[inner sep=0] {\includegraphics[width=0.3\textwidth,trim={8cm 16cm 8cm 7.5cm},clip]{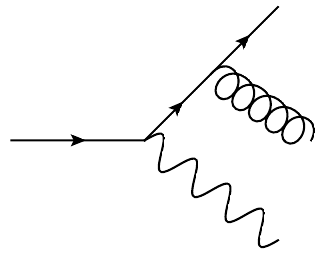}};
        \draw (-1.2, 0.3) node {$t$};
        \draw (1, 1.5) node {$b$};
        \draw (1, -0.1) node {$g$};
        \draw (0.1, -1.1) node {$W$};
    \end{tikzpicture}
    \caption{Illustrative diagram of a first gluon emission from the $b$ quark in a top decay.}
    \label{fig:decaygluon}
\end{figure}

Monte Carlo event generators include phenomenological models of non-perturbative QCD effects which are not necessarily derived from first principles. One such effect that has received a lot of attention recently, is the recoil treatment in top decays. This is an ambiguity that arises in the treatment of the top decays $t \to Wb + X$ in dipole parton showers. The first gluon emission, illustrated in Figure~\ref{fig:decaygluon}, is controlled by matrix element corrections. But for the second gluon emission there are different possible prescriptions for the recoil treatment. Indeed the ATLAS analysis in ref.~\cite{ATLAS:2022jbw} sees an effect of approximately $250$~MeV. 

\begin{figure}[htb]
    \centering
    \begin{subfigure}{.5\textwidth}
        \centering
        \includegraphics[width=\linewidth]{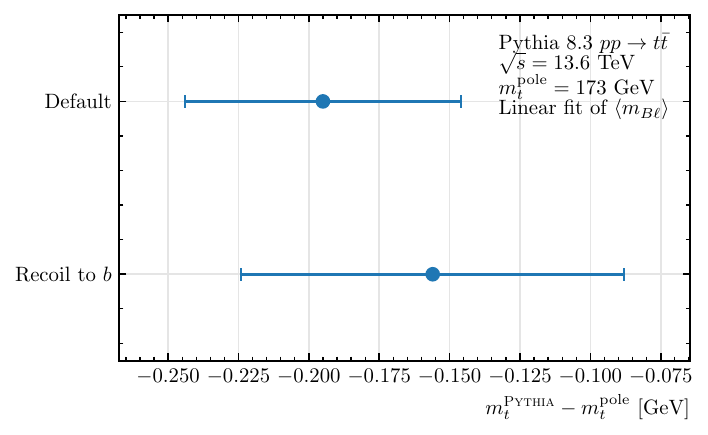}
        \caption{$pp \to t\bar{t}$, at $\sqrt{s} = 13.6$~TeV.}
        \label{fig:mtshiftavepp}
    \end{subfigure}%
    \begin{subfigure}{.5\textwidth}
        \centering
        \includegraphics[width=\linewidth]{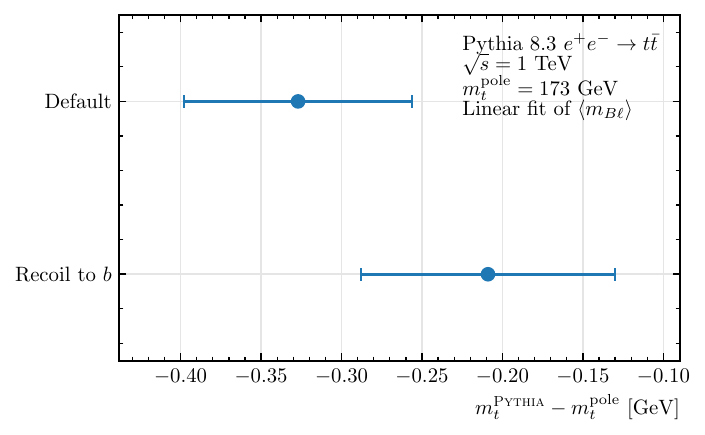}
        \caption{$e^{+}e^{-} \to t\bar{t}$, at $\sqrt{s} = 1$~TeV.}
        \label{fig:mtshiftaveee}
    \end{subfigure}
    \caption{The difference in mass, $m_t^{\pythia{}} - m_t^{\rm pole}$, extracted from the linear fits to $\langle m_{B\ell} \rangle$ shown for default \pythia{} parameters and for the choice of the $b$-quark for the shower recoil in the top
    decays. Adapted from ref.~\cite{Corcella:2026tnu}.}
    \label{fig:mtshiftave}
\end{figure}

In \textsc{Pythia}, there are two options: Letting the $W$ act as the recoiler (the default option) or letting the $b$ quark act as the recoiler. In Figure~\ref{fig:mtshiftave} we show the effect of this option on the linear fit to $\langle m_{B\ell} \rangle$. The effect size this option gives is approximately $40$~MeV for $pp$ collisions and $120$~MeV for $e^{+}e^{-}$ collisions, where the difference is within the respective Monte Carlo errors. 

%%%%%%%%%%%%%%%%%%%%%%%%%%%%%%%%%%%%%%%%%%%%%%%%%

\section{Conclusions}
\label{sec:conclusions}

In this work, we have implemented the formation of heavy-light top mesons in \textsc{Pythia} to understand better the size of effects that uncertainties from non-perturbative modelling in Monte Carlo event generators introduce on the interpretation of the top quark mass. The mass of the top mesons can then be related to the top quark pole mass using Heavy Quark Effective Theory. Then by fitting the input top quark mass in \textsc{Pythia} with the top meson mass, we find a negative mass difference of $240$--$330$~MeV, consistent with $\mathcal{O}(\Lambda_{\rm QCD})$.

For a more detailed discussion, see ref.~\cite{Corcella:2026tnu}.

The implementation of the hadronisation of top quarks into top mesons and their subsequent spectator decay in \textsc{Herwig}~7.3.0 is ongoing and is expected to appear in future work. 

%%%%%%%%%%%%%%%%%%%%%%%%%%%%%%%%%%%%%%%%%%%%%%%%%

\begin{acknowledgement}
The presented work was carried out in collaboration with Gennaro Corcella at the Laboratori Nazionali di Frascati. 
\end{acknowledgement}

%%%%%%%%%%%%%%%%%%%%%%%%%%%%%%%%%%%%%%%%%%%%%%%%%

% Bibliography
\bibliography{biblio}

\end{document}